\documentclass[a4paper,fleqn,usenatbib,useAMS]{mnras}

\usepackage{graphicx} 
\usepackage{times}
\def\etal{\it et~al.}

\title[Characterising the nature of Subpulse Drifting]{Characterising the nature of Subpulse Drifting in Pulsars}
\author[Basu \& Mitra]{Rahul Basu$^{1}$, Dipanjan Mitra$^{2,3}$ \\
$^{1}$ Inter-University Centre for Astronomy and Astrophysics, Pune, 411007, India; rahulbasu.astro@gmail.com \\
$^{2}$ National Centre for Radio Astrophysics, Tata Institute of Fundamental Research, Pune 411007, India \\
$^{3}$ Janusz Gil Institute of Astronomy, University of Zielona G\'ora, ul. Szafrana 2, 65-516 Zielona G\'ora, Poland \\
}
\begin{document}

%\date{Accepted\ldots Received\ldots ; in original form\ldots}

%\pagerange{\pageref{firstpage}--\pageref{lastpage}} \pubyear{2017}

\maketitle

\label{firstpage}

\begin{abstract}
We report a detailed study of subpulse drifting in four long period pulsars. 
These pulsars were observed in the Meterwavelength Single-pulse Polarimetric 
Emission Survey and the presence of phase modulated subpulse drifting was 
reported in each case. We have carried out longer duration and more sensitive 
observations lasting 7000-12000 periods, between frequency range of 306 and 339
MHz. The drifting features were characterised in great detail including the 
phase variations across the pulse window. In two pulsars J0820$-$1350 and 
J1720$-$2933 the phases changed steadily across the pulse window. The pulsar 
J1034$-$3224 has five components. The leading component was very weak and was 
barely detectable in our observations. The four trailing components showed the 
presence of subpulse drifting. The phase variations changed in alternate 
components with a reversal in the sign of the gradient. This phenomenon is 
known as bi-drifting. The pulsar J1555$-$3134 showed the presence of two 
distinct peak frequencies of comparable strengths in the fluctuation spectrum. 
The two peaks did not appear to be harmonically related and were most likely a 
result of different physical processes. Additionally, the long observations 
enabled us to explore the temporal variations of the drifting features. The 
subpulse drifting was largely constant with time but small fluctuations around 
a mean value was seen.
\end{abstract}


\section{Introduction}
\noindent
The radio emission from pulsars is highly pulsed and usually occupy a small 
fraction of the pulsar period. The single pulses consist of one or more 
individual subpulses. In certain pulsars it is seen that the subpulses exhibit 
a periodic variation within the pulse window either in their location or 
amplitude or both and the phenomenon is known as subpulse drifting 
\citep{dra68}. There are around 120 pulsars where this phenomenon has been 
reported \citep{wel06,wel07,bas16}. The subpulse drifting can be broadly 
categorised into two distinct classes, the phase modulated drifting, where the 
subpulses show a steady shift in position across the pulse window, and the 
amplitude modulated drifting, where the subpulses are stationary within the 
pulse window but periodically change in intensity. The fluctuation spectral 
analysis \citep{bac73,bac75}, which involve carrying out Fourier transforms for 
a number of consecutive single pulses along specific pulse longitudes, 
illustrate the different drifting classes in pulsars. The drifting periodicity 
is seen as a peak in the fluctuation spectrum. The subpulse motion across the 
pulse window, on the other hand, is seen as a phase variation corresponding to 
the peak frequency. The amplitude modulated drifting shows a relatively flat 
phase behaviour across the pulse window while the phase modulated drifting 
shows large systematic phase changes. 

The radio emission from pulsars is expected to arise due to the growth of 
instabilities in the relativistic plasma outflowing along the open magnetic 
field lines \citep{ass98,mel00,gil04,mit09}. The plasma has been proposed to 
originate from an inner acceleration region (IAR) above the polar cap in the
form of sparking discharges \citep{rud75}. In this scenario the subpulse 
drifting is related to the plasma dynamics which is governed by the 
$\bf{E}\times\bf{B}$ drift in the IAR. The subpulse drifting is a consequence 
of the sparks lagging behind corotation of the star. Using this concept a new 
relationship between the drifting and the spin down energy loss ($\dot{E}$) was
established by \citet{bas16}. The drifting periodicity ($P_3$), for the phase 
modulated drifting, was found to be anti-correlated with $\dot{E}$, $P_3$ 
$\propto$ $\dot{E}^{-0.6} P$, where $P$ is the pulsar period. The amplitude 
modulated drifting, on the other hand, showed no clear dependence on $\dot{E}$.
Additionally, the phase modulated drifting was only seen in pulsars with 
$\dot{E} <$ 2$\times$10$^{32}$ erg~s$^{-1}$. The periodic subpulse modulation 
seen in pulsars with $\dot{E} >$ 2$\times$10$^{32}$ erg~s$^{-1}$ exclusively 
belonged to the amplitude modulation class. It has been shown by \citet{bas17} 
that the periodicities of amplitude modulation are similar to periodic nulling,
implying a common origin for both. It was also suggested that periodic 
amplitude modulation is a different phenomenon from phase modulated subpulse 
drifting, hereafter simply subpulse drifting. 

As the pulsar rotates the line of sight traverses from the leading to the 
trailing edge of the pulse window. The phase variations, seen from the leading 
to the trailing edge of the window, can be broadly categorised into two 
classes. The first group is called negative drifting and the phases show a 
positive slope from the leading to the trailing edge. The subpulses in negative
drifting are shifted towards leading part of the profile with increasing time. 
The positive drifting on the other hand corresponds to the case where the phase
changes show a negative slope from the leading to the trailing edge of the 
pulse window. In this case the subpulses shift towards the trailing part of the 
profile in subsequent periods. The positive and negative drifting originate due
to different reasons in competing models explaining this phenomenon. According 
to the carousel model \citep{gil00,des01} the subpulse drifting originates due 
to the rotation of the sparking discharges in the IAR around the magnetic axis.
The drift direction can be attributed to either the clockwise or anti-clockwise
rotation of the sparking system. Additionally, depending on whether the line of
sight cuts the emission beam inwards or outwards from the rotation axis the 
drift direction is reversed. In a second model for subpulse drifting, where the 
sparks in the IAR are expected to lag behind corotation \citep{bas16}, the 
drift direction is representative of the aliasing effect around 2$P$. The 
subpulses in subsequent periods have an intrinsic motion from the trailing to 
the leading edge. If the drift periodicity is more than 2$P$ the positive 
drifting is seen. The negative drifting occurs if the drift periodicity is less
than 2$P$.

The Meterwavelength Single-pulse Polarimetric Emission Survey \citep[MSPES,
][]{mit16} was carried out to study the radio emission properties of 123 
pulsars. \citet{bas16} conducted fluctuation spectral studies to characterise 
the periodic subpulse behaviour in these pulsars. They concentrated primarily 
on the drifting periodicity and did not address phase changes associated with 
them in detail. One primary reason for this is such studies require very 
sensitive observations covering a large number of single pulses which are 
restricted in the survey setup. In this work we have characterised the subpulse
drifting in four pulsars observed in MSPES using more sensitive studies. These 
pulsars were selected based on their prominent drifting features as well as 
high signal to noise detections of the drifting peaks in the fluctuation 
spectra. We have carried out long observations of these sources and carried out
the most sensitive phase variation studies for their drifting features. 
In addition, we have also investigated the temporal variations of the subpulse
drifting phenomenon using the long observations.

\section{Observations and Analysis}
\begin{figure*}
\begin{tabular}{@{}cr@{}cr@{}cr@{}}
{\mbox{\includegraphics[scale=0.4,angle=0.]{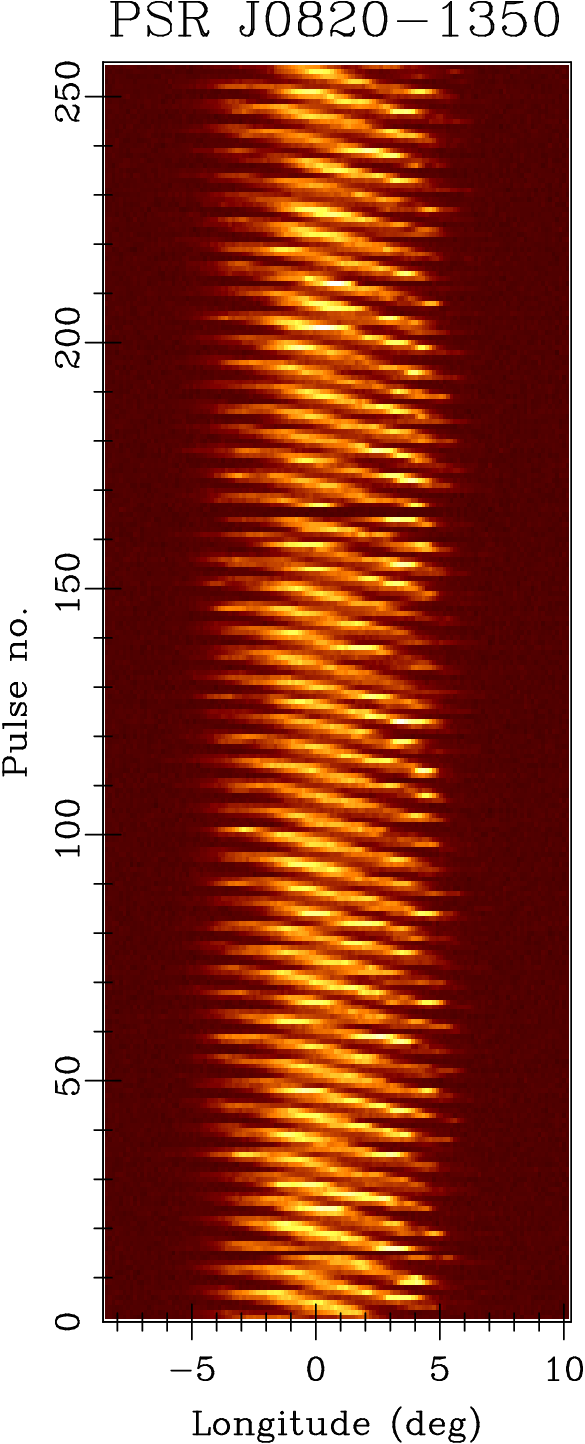}}} &
\hspace{5px}
{\mbox{\includegraphics[scale=0.4,angle=0.]{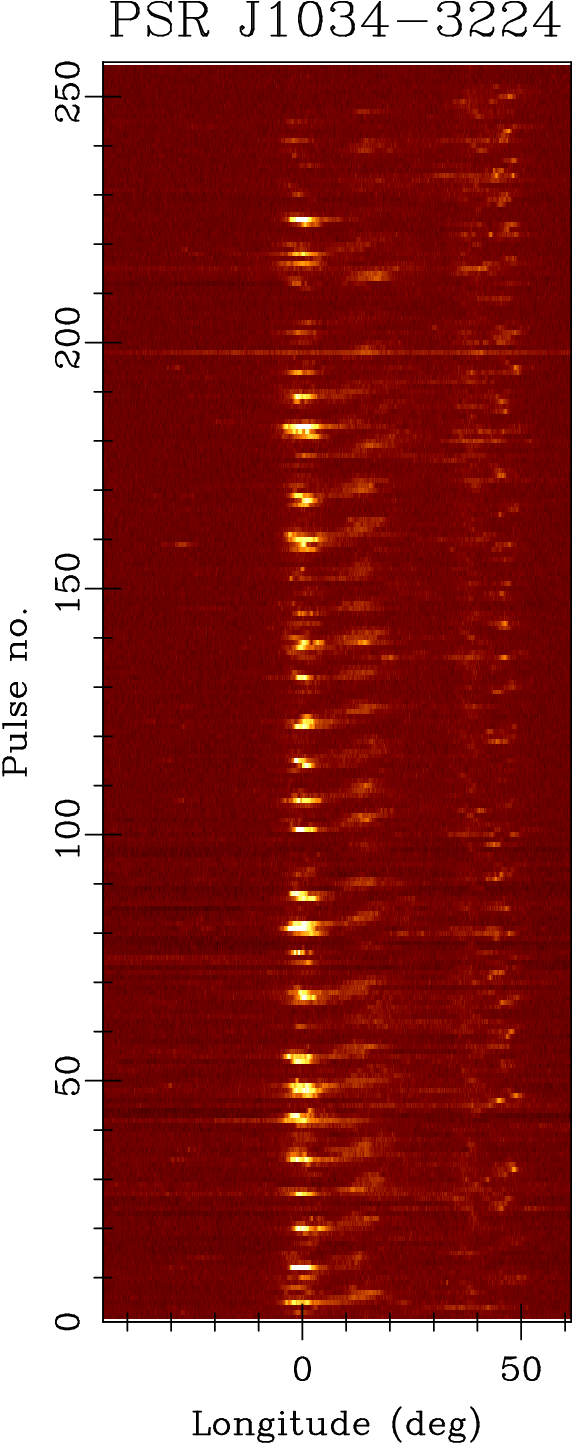}}} &
\hspace{15px}
{\mbox{\includegraphics[scale=0.4,angle=0.]{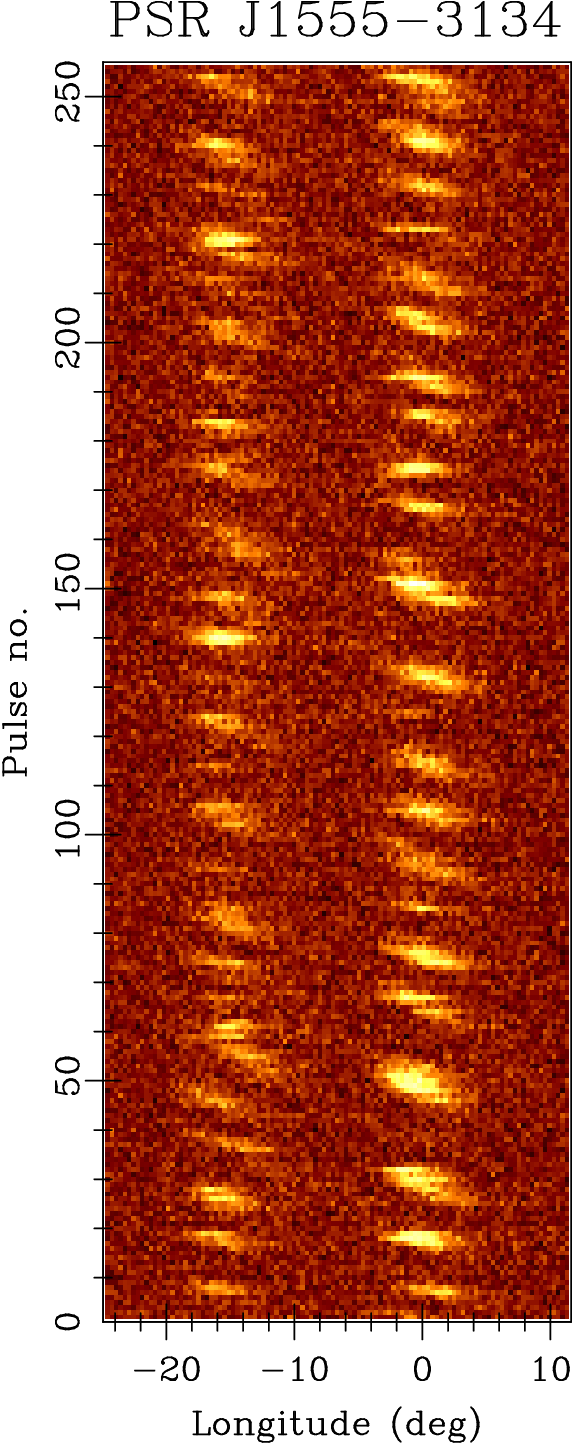}}} &
\hspace{5px}
{\mbox{\includegraphics[scale=0.4,angle=0.]{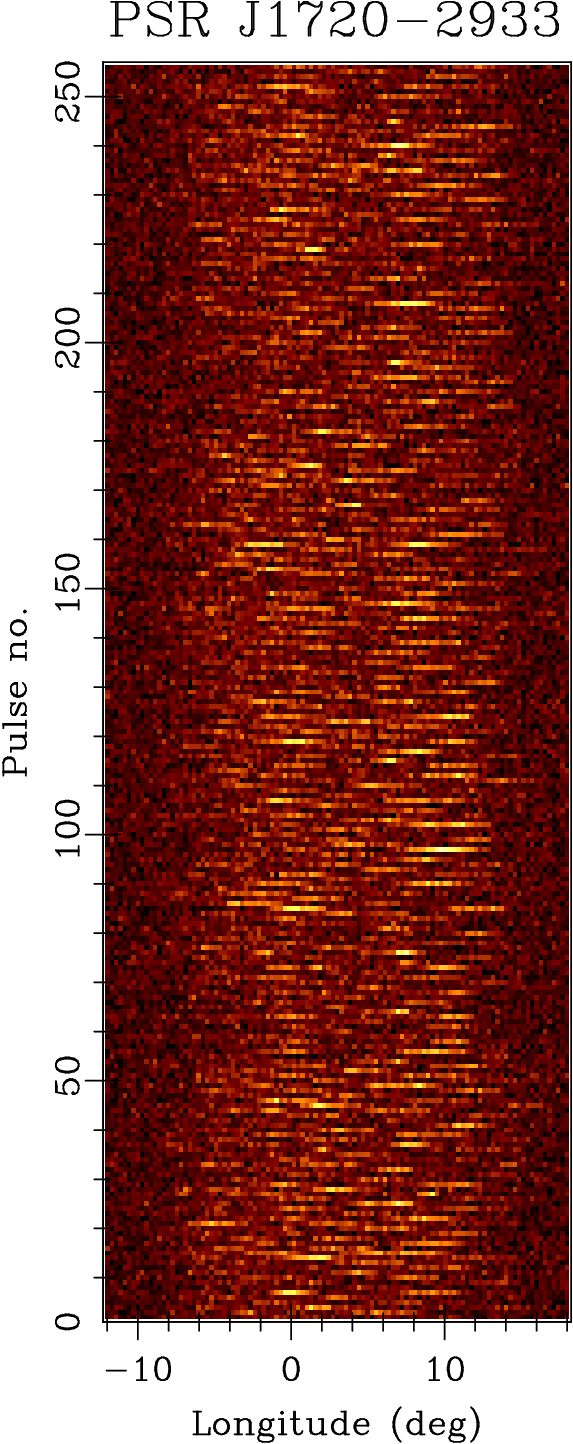}}} \\
\end{tabular}
\caption{The figure shows a section of the single pulse stacks comprising of 
256 consecutive periods for the four the pulsars J0820$-$1350, J1034$-$3224, 
J1555$-$3134 and J1720$-$2933.}
\label{fig_sngl}
\end{figure*}

We have observed the four pulsars in Table~\ref{tabobs} using the Giant 
Meterwave Radio Telescope (GMRT), which is located near Pune, India 
\citep{swa91}. The GMRT is an interferometric array consisting of thirty 
antennas each of forty five meters diameter, with fourteen antennas located 
within a central square kilometer area and the remaining sixteen antennas 
spread along three arms in a Y-shaped array. We have used the Telescope in the 
phased array mode where the signals from different antennas were co-added. In 
order to reach sufficient sensitivity for single pulse studies we used 
approximately twenty antennas, including all the available central square 
antennas and the two nearest arm antennas. A phase calibrator was recorded at 
the start of each observation and subsequently after every hour. Appropriate 
``phasing'' solutions were estimated to correct for temporal gain variations 
in each antenna. The pulsars were observed on November 4 and 5, 2015 with each 
pulsar recorded for durations between 7000 and 12000 periods. In some cases 
there were phasing breaks in between the observations of a source.

Total intensity signals were recorded at frequencies between 306 and 339 MHz,
covering a bandwidth of 33 MHz and spread out over 256 frequency channels. This 
resulted in higher sensitivity detection of single pulses compared to MSPES, 
where the signals were recorded in the full polarization mode but over 16 MHz 
bandwidth. The time resolution for these observations was 491.52 microseconds. 
The dispersion spreads across the frequency band were corrected using the known 
dispersion measures of the pulsars and the signal was subsequently averaged 
along the frequency band to produce a time series corresponding to the radio
emission from each pulsar. In order to maintain continuity in the overall time 
series data during phasing breaks suitably weighted noise signals were inserted
during these intervals. Finally, the time series signals were re-sampled to 
form a two dimensional pulse stack with one axis along the pulse longitude, 
separated into integral bins, and the other along the pulse number (see figure 
\ref{fig_sngl}). Further information about the initial analysis is detailed in 
\citet[][see appendix A]{bas16}.

We have used the fluctuation spectral analysis to explore the subpulse 
drifting in the four pulsars studied in this work \citep{bac73,bac75}. The 
primary analysis scheme was the Longitude Resolved Fluctuation spectra (LRFS) 
where Fourier transforms across each longitude, for 256 consecutive single 
pulses, were carried out. Subsequently, the starting point was shifted by fifty
periods and the process was repeated till the end of the observations. Each 
LRFS realisation had two constituents, the amplitude with one or more peak 
frequencies ($f_p$ in units of cycles/$P$), and the corresponding phase at 
$f_p$ giving the nature of subpulse variation across the pulse window. In 
\citet{bas16} we proposed a method to quantify the temporal variation in LRFS 
amplitude by stacking the average across the pulse window as a function of the 
starting period. Two consecutive spectra separated in this work were separated
by ten periods\footnote{we have replaced ten periods by fifty periods in this 
work. We found this did not affect the estimation of temporal variation of the 
signal, since, this was well below the FFT lengths, but this considerably saved 
computation time.}). In order to reduce the effect of the unknown baseline 
level the time average peak intensity in the LRFS was normalized to unity. The 
primary drawback of this method was that the information across the the pulse 
window was lost due to averaging. Also, no information was preserved regarding 
the phase variations. Additionally, it was seen that the gaps in the data 
during telescope phasing resulted in strong low frequency peaks which 
suppressed the drifting peaks in the average spectra. In order to mask these 
unwanted peaks we estimated the baseline rms level of the LRFS and replaced the
affected LRFS intervals during phasing with simulated noise.

\begin{table}
\resizebox{\hsize}{!}{
\begin{minipage}{80mm}
\caption{Observing Details.}
\centering
\begin{tabular}{ccccc}
\hline
% & & \\
 Pulsar & Period & DM  & $\dot{E}$ & Npulse \\
  & (sec) & (pc~cm$^{-3}$) & (10$^{31}$erg~s$^{-1}$) &  \\
\hline
  &  &  &  &  \\
 J0820$-$1350 & 1.238 & ~40.94 &  4.38~ & 10199 \\
  &  &  &  &  \\
 J1034$-$3224 & 1.150 & ~50.75 &  0.597 & ~7837 \\
  &  &  &  &  \\
 J1555$-$3134 & 0.518 & ~73.05 &  1.77~ & ~6947 \\
  &  &  &  &  \\
 J1720$-$2933 & 0.620 & ~42.64 & 12.3~~ & 11660 \\
  &  &  &  &  \\
\hline
\end{tabular}
\label{tabobs}
\end{minipage}
}
\end{table}

In this work we have devised another method to study the nature of subpulse 
drifting using the fluctuation spectra. Similar, to the previous case the 
individual FFTs for 256 consecutive periods were estimated for the entire 
observing run by shifting the starting point by fifty periods. The first 
step was to estimate the peak amplitude of the fluctuation spectra as a 
function of the pulse longitude. The $f_p$ from the average LRFS and the 
corresponding error, $\delta f$, was determined. Here $\delta f$ = 
FWHM/2.355, FWHM being the Full Width at Half Maximum of the peak 
\citep{bas16}. Now, for each 256 period FFT we looked for the maximum value 
within the error window at each longitude (i.e $f_p-3\delta f < f < f_p+3\delta
f$) and any significant measurement ( $>$ 3 times the rms level of the 
baseline) was identified. In addition the $f_p$s at each longitude, for all
the different measurements of the LRFS, were averaged and the value was 
overlayed on top of the individual measurements. This is shown in Figure 
\ref{fig_J0820} (right plot, top panel), where the small red dots represent the
maximum amplitude corresponding to $f_p$ for all LRFS realization and the 
black dots show the average value. The phase variations, corresponding to the 
peak amplitude, were also estimated similarly (see Figure~\ref{fig_J0820}, 
right plot, middle panel). However, one additional step was followed while 
estimating the phase variations. The fifty period shift between two consecutive
LRFS was not an integral multiple of the drifting periodicity ($P_3$) and 
resulted in arbitrary phase differences between different realizations of the 
LRFS. This would result in smearing of the peak phase information in the 
average behaviour. To address this issue we have a priori fixed the phase at a 
certain pulse longitude, the profile peak intensity, to be zero and 
estimated the variation of the phases across the window. The absolute phase 
information during each LRFS is lost but the information regarding the 
relative phase variation across the pulse window is not washed away. For both 
the estimates of the peak amplitude and phase at any longitude we have only 
considered significant measurements, i.e. the measured frequency peak was 
greater than 3$\sigma$ level of the baseline. 

The above method of estimating the variation of drift phase is a superior 
technique for pulsars with long uninterrupted drifting patterns. This is 
primarily because some of the phase information at certain longitudes which 
were lost during any specific LRFS realization due to weaker signal strengths 
can be recovered during the longer integrations. Earlier techniques \citep[eg.
][]{van02} have used prominent single pulse sequences to establish the nature 
of subpulse motion within the pulse window. In these cases the results were 
likely to be biased by that particular sequence of single pulses. The 
sensitivity of estimating the subpulse motion would also be affected by the 
errors in localising the subpulse peaks. We have also determined the average 
pulsar profile within the pulse window as shown in right plot, bottom panel of 
figure~\ref{fig_J0820}. We did not have absolute flux calibration for the 
profiles and normalized the profile peak to unity in these plots.

\section{The Subpulse Drifting in Individual Pulsars}
\noindent
We have applied the analysis schemes described in the previous section to study
in detail the temporal as well as longitudinal variation of subpulse drifting
in all four pulsars. We have employed two main analysis techniques for these 
studies.The first technique involved averaging the LRFS across the pulse 
window and estimating its temporal variation. The second technique 
involved finding the peak amplitude and look for its variation, as well as the
phase associated with it, as a function of longitude. All the time realizations
in the second exercise were seen as a spread around a mean value at each 
longitude.

\subsection{J0820$-$1350}
\begin{figure*}
\begin{tabular}{@{}lr@{}}
{\mbox{\includegraphics[scale=0.5,angle=0.]{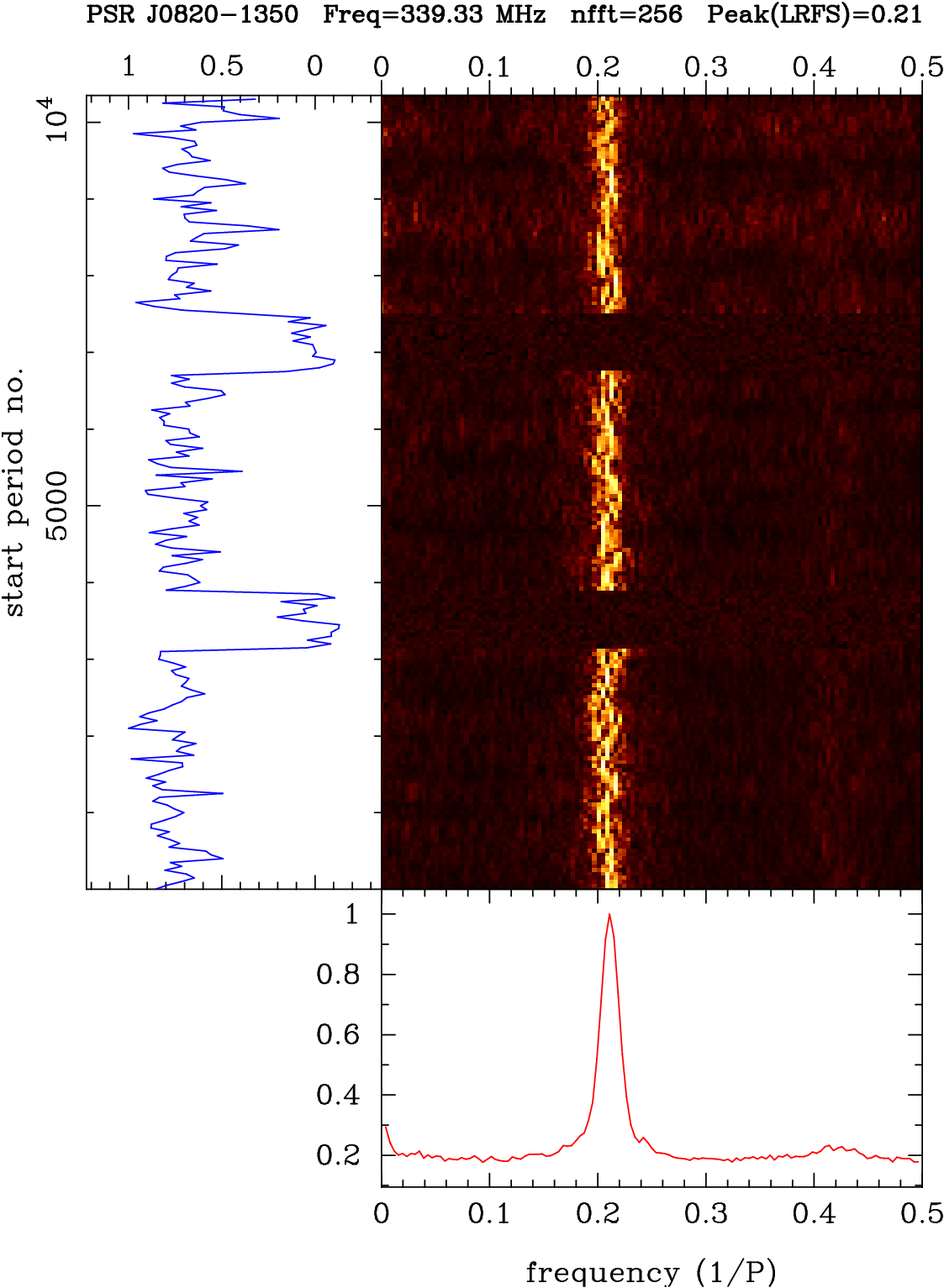}}} &
{\mbox{\includegraphics[scale=0.5,angle=0.]{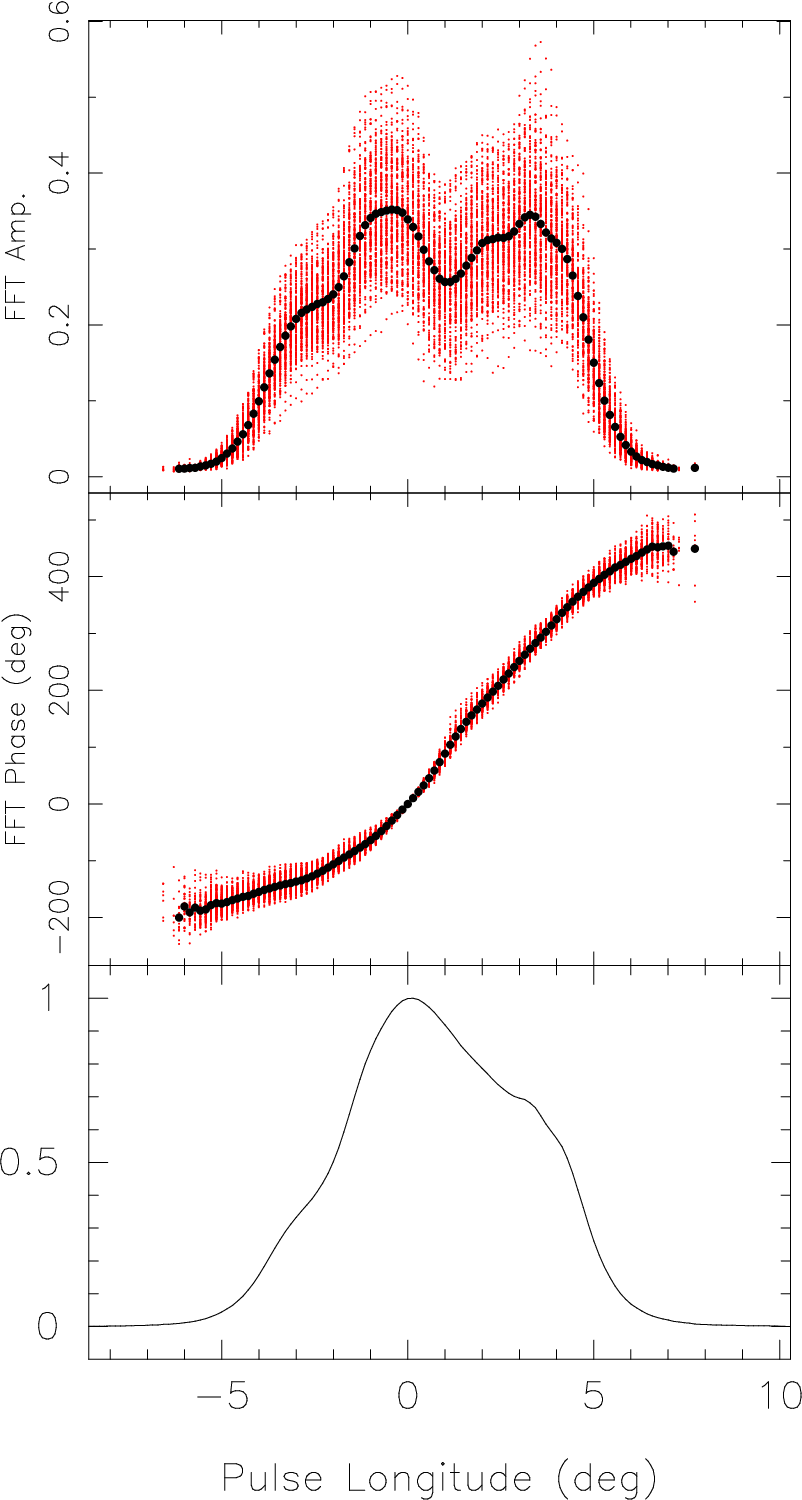}}} \\
\end{tabular}
\caption{The figure shows the temporal variation of the LRFS in the pulsar 
J0820$-$1350. Each LRFS was determined for 256 consecutive periods and the 
temporal variations were estimated by shifting the starting period by fifty 
pulses. The left plot shows the average LRFS across the whole pulse window as a 
function of the starting period. The two regions in between with no signal 
corresponds to breaks in the observations during phasing of antennas which have
been masked with simulated noise. The right panel represents the variation of 
the LRFS across the pulse window for different time realizations for the peak
amplitude (top panel), the corresponding phase variations (middle panel) and 
the average profile (bottom panel). The temporal variations are seen as a 
spread of the amplitudes as well as the phases while the average value at each 
longitude is shown as a black dot. The profile peak was the reference point for
aligning the phases in each LRFS.}
\label{fig_J0820}
\end{figure*}

In figure \ref{fig_J0820}, left plot, the temporal variation of the average 
LRFS is shown. The peak frequency is $f_p$ = 0.211$\pm$0.009 cycles/$P$ and 
the Full Width Half Maximum (FWHM) of the peak is 0.021 cycles/$P$. This 
corresponds to drifting periodicity $P_3$ = 4.7$\pm$0.2 $P$. In order to 
estimate the relative strength and spread of the drifting feature an $S$ factor
was introduced by \citet{bas16}. This was defined as $S$ = $V_p$/FWHM, where 
$V_p$ is the separation between the frequency peak and the baseline level of 
the fluctuation spectra. The time averaged LRFS peak for this measurement had 
$S$ = 38.9. The above results for peak in the LRFS and the corresponding 
drifting periodicity is compatible with the measurements reported in 
\citet{bas16}. The frequency peak despite its relatively high $S$ value is not 
sharply defined either in the individual LRFS or in their time evolution. The 
spread of the peak frequency in the LRFS has interesting implications about the
physical conditions in the IAR which we discuss in the next section. The right 
plot in figure \ref{fig_J0820} presents the change of drifting features across 
the pulse window. The top part shows the variation of the peak intensity while 
the middle part shows the variation of the phase associated with the peak. The 
pulse window, defined at 5$\sigma$ above baseline level, is between 
-8.5\degr~and +10\degr~longitude with the peak intensity at 0\degr. A dip in 
the LRFS peak amplitude is seen around +1\degr~from the intensity peak while 
the corresponding peaks on either side were at -0.5\degr~and +3.0\degr. The 
peak phases (middle panel, right plot) show a large variation ($>$ 600\degr) 
across the pulse window. The phases across the pulse window do not vary in a 
linear manner. The phase variations are flatter in the leading part of the 
profile which becomes more steep around the pulse peak and then once again 
relatively flat near the trailing edge. 

The pulsar J0820$-$1350 is a well known example of nulling and drifting 
appearing in the same system. The subpulse drifting was first measured by 
\citet{lyn83} who reported a change in the drift rate before and after the 
onset of nulling. We have detected short duration nulls lasting between 1-2 
periods. No periodicity was detected associated with the occurrence of these 
short nulls \citep{bas17}. The change in drifting behaviour around the 
onset of nulls has been reported in other pulsars like B0031$-$07 
\citep{viv97,smi05,mcs17} and B0809+74 \citep{lyn83,van02}. However, the state 
changes are more distinct in these pulsars and nulling lasts for longer 
durations which makes such identification more easier. In fluctuation spectral 
analysis any deviation in subpulse tracks would show up as deviations in the 
phase behaviours. As seen in the phase variations for the entire observing run 
the phases are bunched up together indicating no large jumps or variations 
around the onset of short nulls. However, we cannot rule out small phase 
variations around nulls which can be hidden within the spread of the phase 
measurements. The median value of the spread in the phase bunches across all 
longitudes is 11.9$\pm$2.4\degr. A detailed study of the subpulse variation 
across the pulse window was carried out by \citet{big87}. Their estimates of 
the variation of the amplitude of the peak frequency across the pulse window is
consistent with our results. They estimated the phase difference between
two longitude bins and showed the phase track to be steeper near the center of 
the profile and relatively flatter near the edges. This is consistent with our 
measurements. The subpulse drifting in this pulsar has also been studied by
\citet{wel06,wel07} at two frequencies, 325 and 1360 MHz. They used the two 
dimensional fluctuation spectrum \citep[2DFS,][]{edw02} and reported horizontal 
structures which is indicative of non-linear drift bands at both wavelengths. 
The relative steepening of the phase variations towards the center of the 
profile was also seen in their work with the effect being more pronounced at 
the higher frequency.
\citet{jan04} using the carousel model suggested that the drifting slows down 
during the null state and the subpulse drift in this pulsar was aliased. The 
pulsar has also been detected in intermittent and longer null states in the 
past. Despite our relatively long observations the emission was in the bright 
state throughout. However, the 333 MHz observations in MSPES found the pulsar 
in the low intensity/null states. The pulsar in this state showed the presence 
of drifting with exactly the same properties as the bright state.

\subsection{J1034$-$3224}
\begin{figure*}
\begin{tabular}{@{}lr@{}}
{\mbox{\includegraphics[scale=0.5,angle=0.]{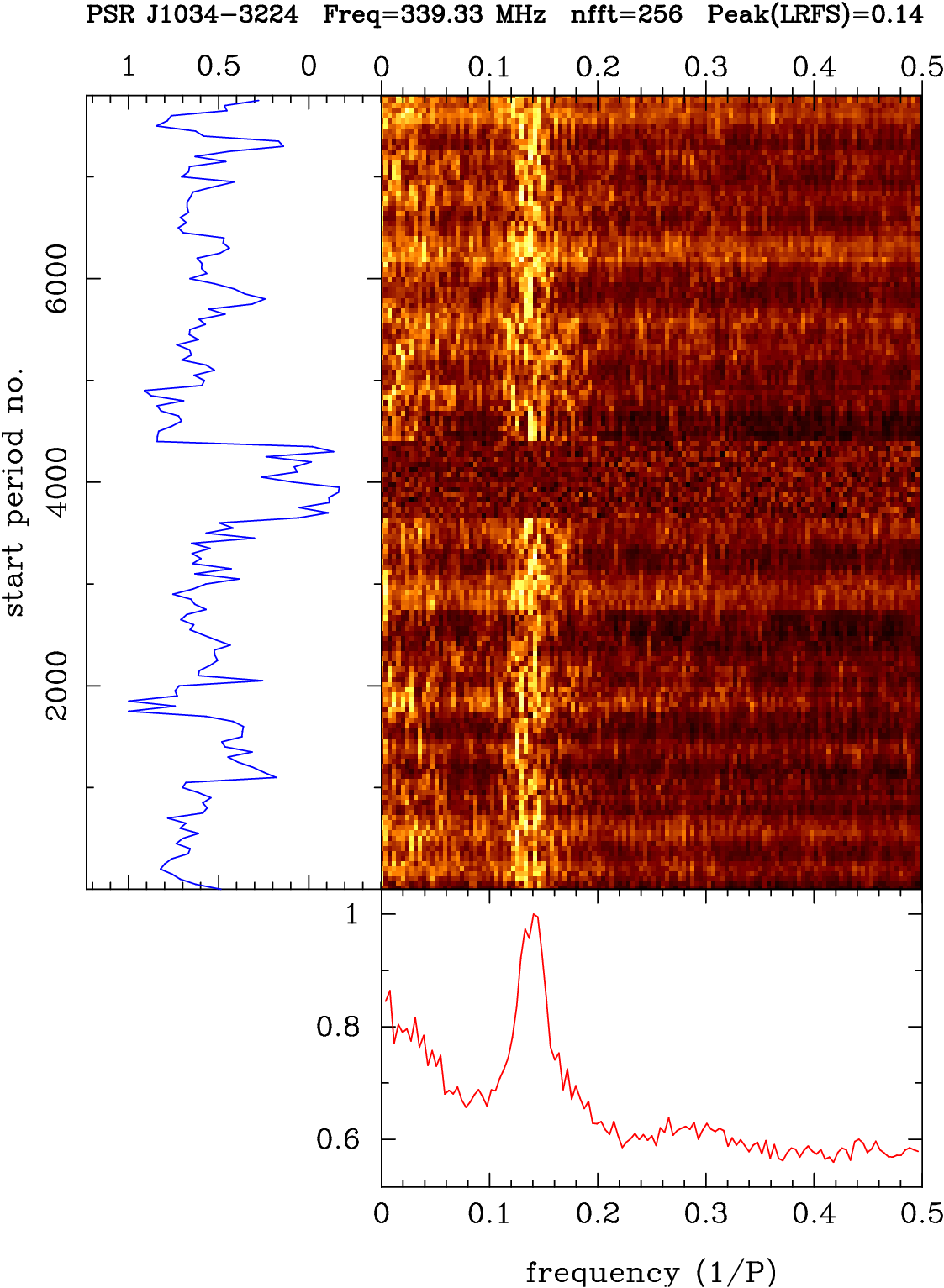}}} &
{\mbox{\includegraphics[scale=0.5,angle=0.]{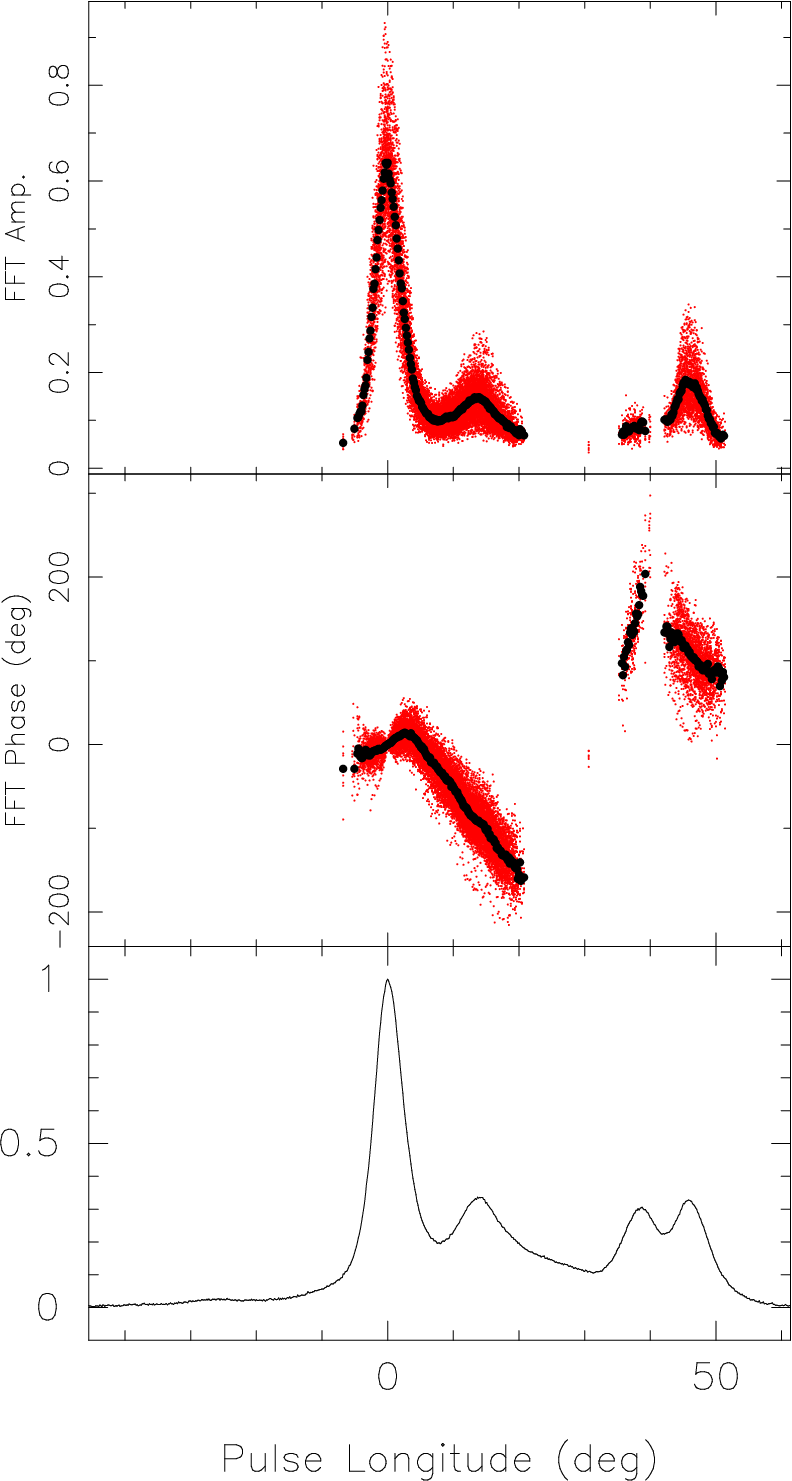}}} \\
\end{tabular}
\caption{The figure shows the temporal variation of the LRFS in the pulsar 
J1034$-$3224. Each LRFS was determined for 256 consecutive periods and the 
temporal variations were estimated by shifting the starting period by fifty 
pulses. The left plot represents the average LRFS across the whole pulse window
as a function of the starting period. The region in between with no signal 
corresponds to breaks in the observations during phasing of antennas which have
been masked with simulated noise. The right plot shows another realization of 
the LRFS by showing the variations across the pulse window of the peak 
amplitudes for all times (top panel), the corresponding phase variations (middle
panel) along with the average profile (bottom panel). The temporal variations 
are seen as a spread of the amplitudes as well as the phases while the average 
value at each longitude is shown as a black dot. The profile peak was the 
reference point for aligning the phases in each LRFS.}
\label{fig_J1034}
\end{figure*}

The pulsar with five components has a relatively wide profile with the 
5$\sigma$ boundaries lying between -45\degr~and +61\degr, where the peak 
intensity is centered around zero longitude. The leading component is very weak
and appears sporadically with the peak located around -28\degr~and is similar 
to pre-cursor emission \citep{bas15}. The subpulse drifting in this pulsar was 
reported for the first time by \citet{bas16}. No drifting feature could be 
detected for the leading component. The remaining four components all showed 
subpulse drifting with the same periodicity as shown in figure \ref{fig_J1034}.
The left plot of the figure shows the temporal variation of the LRFS where the 
peak frequency is $f_p$ = 0.139$\pm$0.014 cycles/$P$. This corresponds to a 
drift periodicity of $P_3$ = 7.2$\pm$0.7. The FWHM of the peak is 0.033 
cycles/$P$ which puts the strength of the peak value to be $S$ = 12.6. These 
measurements of the drifting properties are consistent with previous values. 
The temporal variations of the LRFS shows that the peak frequency has a certain
width around the peak value which has interesting implications for the IAR. 
Additionally, the time average LRFS shows the presence of a wider structure in 
the fluctuation spectra peaking at zero frequency and sitting underneath the 
subpulse drifting peak. The origin of this wider structure in the fluctuation 
spectra was not clear. There are presence of horizontal stripes seen in the 
time variation of the LRFS. These stripes are indicative of the 
alteration in the baseline level of the fluctuation spectra with time. The 
baseline level corresponds to the mean value of the signal and would be zero if
the mean value is zero. In our case the existence of horizontal stripes imply 
that the mean profile value over 256 periods is not constant but changes with 
time. If we had a continuous shift in the starting period rather than a 50 
period shift the variations would be continuous and not appear as stripes. The 
stripes are most visible in this pulsar due to its relatively weaker peak 
amplitude which makes the baseline variation more clearly visible.

The right plot in figure \ref{fig_J1034} shows the variation of the peak 
amplitude (top panel) as well as the phases associated with them (middle panel) 
as a function of the pulse longitude. The amplitude of the peak frequency
is maximal for the second component which is also the strongest in the 
profile. The third and fifth components are relatively weaker with roughly one 
fourth of the second component while the fourth component has weakest drifting 
with about one tenth the peak value despite it being of comparable strength in 
the profile with the third and fifth component. The phase variation on the 
other hand shows the rarest and fascinating phenomenon of bi-drifting 
\citep{qia04b}. The phase variations in the second and fourth components have 
positive slope while the third and fifth components have negative slope, 
indicating drift reversals between adjacent components. The improved phase 
estimation technique in this work have made the identification of bi-drifting 
possible in this pulsar which was earlier proclaimed to have a complex phase 
behaviour \citep{bas16}. The analysis technique loses the absolute phase 
information of the peak frequency, however the relative phase information 
across the profile is preserved. We have measured the phase variations in 
different parts of the profile which apart from the reversals are more or less 
linear. The absolute phase relationship between the leading and trailing 
part of the profile could not be measured due to the gap in between. Thus a 
possible phase shift of 360\degr~may be present between these two parts of the 
pulse window. The phase roughly increases from -16\degr~to +12\degr~in the 
longitude range -4\degr~to +3\degr~corresponding to the second component. The 
third component is wider and spread between the longitude range +3\degr~to 
+20\degr. The phase in this region roughly decreases from +12\degr~to 
-150\degr. The drifting in the fourth component with the weakest features is 
confined between longitude range +35\degr~and +39\degr and the phase increases 
from +70\degr~to +200\degr. Finally, the fifth component has drifting features 
extending from +42\degr~to +52\degr~in longitude with the phase decreasing from
+150\degr~to +80\degr.

As mentioned earlier, the bi-drifting phenomenon is extremely rare and reported
in two other pulsars, J0815+0939 \citep{cha05,sza17} and B1839$-$04 
\citep{wel16}. In case of PSR J0815+0939 the pulsar exhibit four components 
with the opposite phase variation seen only in the second component. In case of
B1839$-$04 there are two components whose phases show opposite slopes. In this 
regard PSR J1034$-$3224 is unique since it shows multiple drift reversals with 
alternate components having opposite direction of phase variations. The 
bi-drifting phenomenon is extremely difficult to understand from simple models 
of the IAR which we briefly discuss in the next section.

\subsection{J1555$-$3134}
\begin{figure*}
\begin{tabular}{@{}lr@{}}
{\mbox{\includegraphics[scale=0.5,angle=0.]{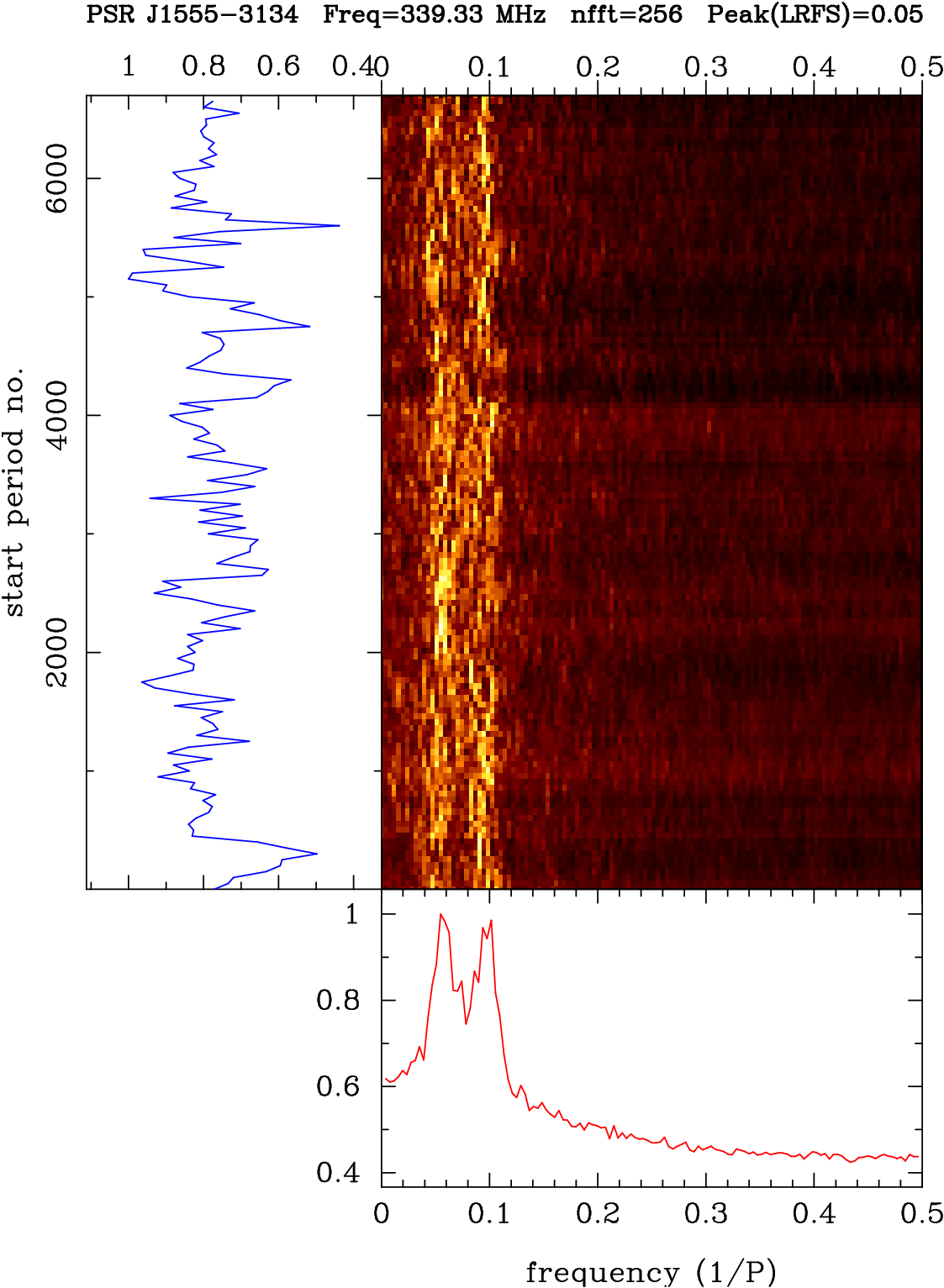}}} &
{\mbox{\includegraphics[scale=0.5,angle=0.]{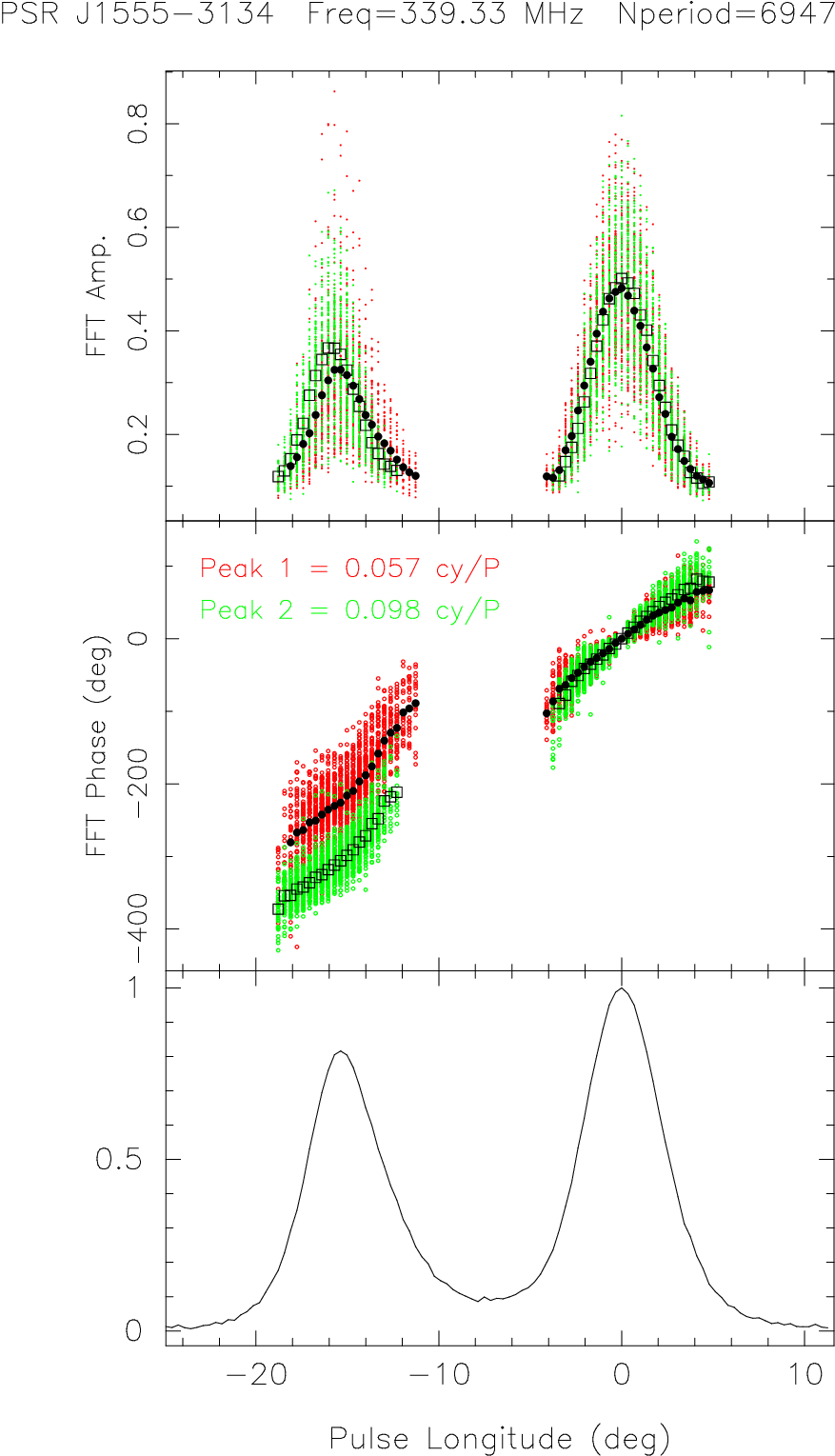}}} \\
\end{tabular}
\caption{The figure shows the temporal variation of the LRFS in the pulsar
J1555$-$3134. Each LRFS was determined for 256 consecutive periods and the
temporal variation was estimated by shifting the starting period by fifty 
pulses. The left plot represents the average LRFS across the whole pulse window
as a function of the starting period. The fluctuation spectra shows the 
presence of two distinct frequency peaks. The right plot shows the variation of 
the LRFS across the pulse window for all times for the peak amplitudes (top 
panel), the corresponding phase variations (middle panel) along with the 
average profile (bottom panel). The temporal variations are seen as a spread of
the amplitudes as well as the phases while the average value at each longitude 
is shown as a black dot. The profile peak is the reference point for aligning 
the phases in each LRFS. The phase variations for the two peaks are shown 
separately and exhibit a difference across the pulse window.}
\label{fig_J1555}
\end{figure*}

The pulsar shows a classical conal double profile with both components showing 
the presence of subpulse drifting \citep{bas16}. The components are of 
comparable intensity with the trailing component associated with the profile
peak. The 5$\sigma$ boundaries of the pulse window are between longitudes 
-24\degr~and +11.5\degr~where the profile peak is aligned along 0\degr. The 
time varying LRFS of the pulsar is shown in figure~\ref{fig_J1555}, left plot, 
where both components have similar drifting periodicities. The pulsar shows the
presence of two distinct peaks in the fluctuation spectra with $f_{p,1}$ = 
0.057$\pm$0.012 cycles/$P$ and $f_{p,2}$ = 0.098$\pm$0.010 cycles/$P$. Though 
the errors are large the peaks do not appear to be harmonically related. The 
corresponding strengths of the peaks are $S_1$ = 20.6 and $S_2$ = 22.2 which 
shows that they are of similar strengths. The FWHM of the peaks are 0.027 
cycles/$P$ and 0.025 cycles/$P$ respectively and the peaks show some spread in 
single realizations of LRFS as well as in their time evolution. The drifting 
periodicities are estimated to be $P_{3,1}$ = 17.5$\pm$3.6 $P$ and $P_{3,2}$ = 
10.2$\pm$1.0 $P$ for the two peaks which is consistent with previous 
measurements.

In figure \ref{fig_J1555}, right plot, the variations of the peak amplitude 
(top panel) as well as the associated phase (middle panel) are shown as a 
function of pulse longitude separately for the two different peaks. 
The peak amplitudes of the two drifting features are not coincident with their 
maximum being slightly displaced along the x-axis (pulse longitude). As a 
first step we estimated the phase variations of the two frequency peaks without
introducing any arbitrary phase shifts across the pulse window. We estimated 
multiple iterations of the 256-period LRFS which will have random phase shifts
between them. This enabled us to test if the two frequency peaks are 
harmonically related. If we assume that the phase variations are linear then 
the first harmonic will show a phase variation with a gradient which is twice 
that of the fundamental frequency. Thus for different iterations of the LRFS 
the phase at any longitude corresponding to two harmonics will have a definite 
relationship. However, the phases corresponding to the two frequency peaks in 
this pulsar had no definite relationship but were randomly distributed. This 
indicates that they are not harmonically related.

In the figure \ref{fig_J1555}, right plot, middle panel, we have treated the 
phases corresponding to the two frequency peaks as independent measurements and
fixed separately to be zero at the longitude of the profile peak. For each
frequency peak the phases across the two components are locked indicating a 
clear phase relationship across the pulse window. The phase variations are 
increasing from the leading to the trailing part of the profile and exhibits a 
positive slope. In order to characterise the slopes of the phase tracks 
for the two frequency peaks we have estimated approximate fits of these tracks 
for the two components using linear least squares. The tracks for each 
256-period LRFS have been fitted separately and the distributions of the slopes
for all such realizations of LRFS were estimated. We have used the original 
phase tracks for this fits without fixing the phase at profile peak to zero. In
Table \ref{tabphsfit} the mean of the slopes for different components and 
frequency peaks as well as the errors (the rms of the distribution) are shown.
It is not possible to directly compare the phase tracks of the two 
frequency peaks since they are not phase locked. But it is interesting to note 
that the slopes are similar for the different peaks which is contrary to a 
harmonical dependence between them.

\begin{table}
\resizebox{\hsize}{!}{
\begin{minipage}{80mm}
\caption{The estimates of the slopes for the different components and frequency
peaks. The phase tracks corresponding to each 256-period LRFS have been 
approximated with linear fits. We report the the mean and rms (error) of the 
distribution of the slopes for the observing durations.}
\centering
\begin{tabular}{cccc}
\hline
  \multicolumn{2}{c}{leading Comp.} & \multicolumn{2}{c}{Trailing Comp.} \\
  Peak 1 & Peak 2 & Peak 1 & Peak 2 \\
\hline  
   &  &  & \\
 22.1$\pm$7.6 & 20.3$\pm$4.7 & 18.5$\pm$3.7 & 21.1$\pm$2.6 \\
% 28.9$\pm$1.0 & 229.5$\pm$15.1 & 18.1$\pm$0.6 & -2.0$\pm$0.9 \\
   &  &  & \\
%Peak 2 & 22.4$\pm$1.0 & 46.7$\pm$16.2 & 20.8$\pm$0.6 & -0.2$\pm$1.0 \\
%  &  &  &  & \\
\hline
\end{tabular}
\label{tabphsfit}
\end{minipage}
}
\end{table}

There is a possibility that the two peaks are not actually simultaneous 
but the pulsar switches very rapidly between the two drifting states at 
timescales less than 256 periods. To test this we used 128 period and 64 period
FFT lengths for the LRFS studies. The two peaks were also seen in these 
measurements indicating their near simultaneity. Any shorter duration studies 
were not possible since the two peaks merged together and could not be 
resolved. There have been earlier observations of systems which undergo mode 
changing with associated change in subpulse drifting periodicity. However, to 
the best of our knowledge PSR J1555$-$3134 is currently the only known example 
where two different drifting periodicities, which are not harmonically related,
exist simultaneously. It is also possible that the pulsar switches between 
these two drifting states at very short timescales. This raises many questions 
about the conditions in the IAR that can accommodate multiple drift 
periodicities either existing simultaneously or switching between them at short
intervals.

\subsection{J1720$-$2933}
\begin{figure*}
\begin{tabular}{@{}lr@{}}
{\mbox{\includegraphics[scale=0.5,angle=0.]{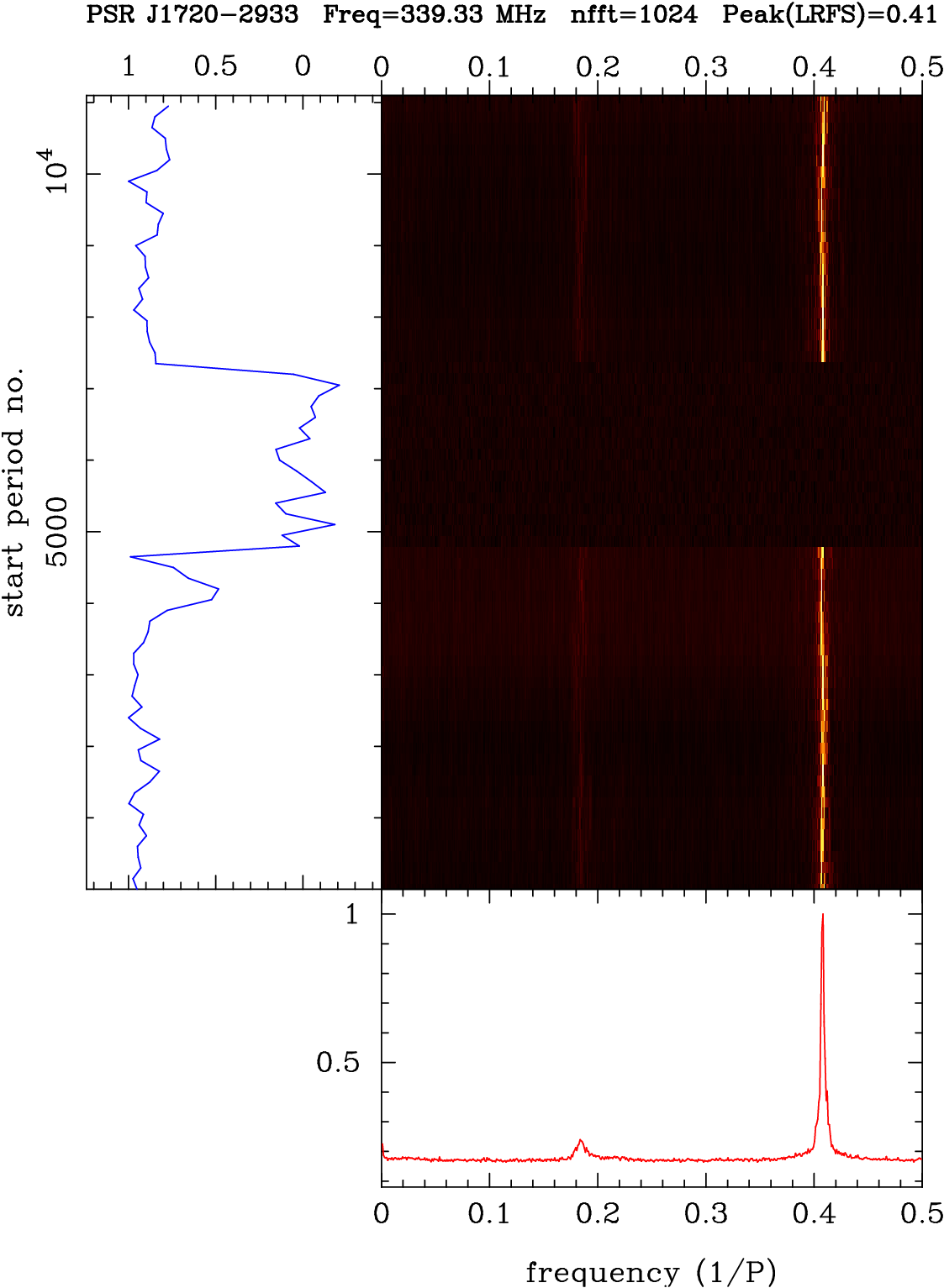}}} &
{\mbox{\includegraphics[scale=0.5,angle=0.]{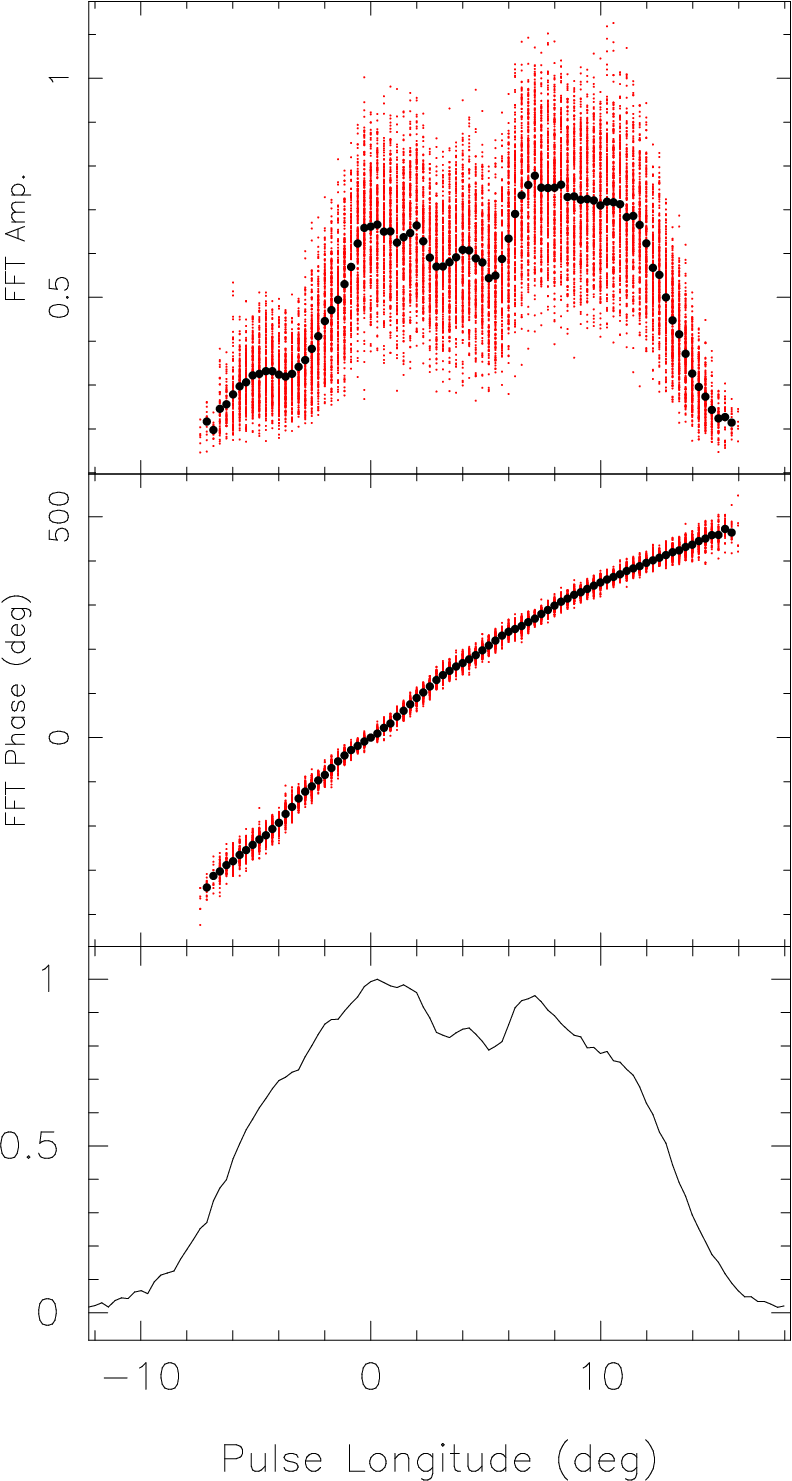}}} \\
\end{tabular}
\caption{The left plot of figure shows the temporal variation of the LRFS in 
the pulsar J1720$-$2933. The pulsar shows one of the narrowest features in the
fluctuation spectra. We have used the a higher resolution studies with FFT 
lengths of 1024 periods to estimate the width of the feature. The temporal 
variation was estimated by shifting the starting period by fifty periods. The 
region in between with no signal corresponds to break in the observations 
during phasing of antennas. The right plot shows the variations of the LRFS 
across the pulse window for all times for the peak amplitude (top panel), the 
corresponding phase variations (middle panel) along with the average profile 
(bottom panel). To obtain higher sensitivity measurements, particularly for the
phase, we have used 256 periods to estimate the LRFS for these studies similar 
to other pulsars. The temporal variations are seen as a spread of the amplitude
as well as the phase while the average value at each longitude is shown as a
black dot.The profile peak was the reference point for aligning the phases in 
each LRFS.}
\label{fig_J1720}
\end{figure*}

The pulsar J1720$-$2933 is another example where the subpulses seem to move 
across the entire pulse window, similar to PSR J0820$-$1350. The pulsed 
emission is roughly spread between longitude range -12\degr~and +18\degr. The 
time variation of the LRFS is shown in figure \ref{fig_J1720}, left plot. 
Unlike the previous cases the drifting feature in the LRFS is much narrower in 
individual realizations and shows a jitter with time resulting in a wider 
average width. To explore the actual width of the drifting feature we have 
carried out higher resolution analysis. This involved using increasing lengths 
of pulse sequences to estimate the fluctuation spectra. It was seen that the 
width stabilized between FFT lengths of 1024 and 2048 periods. We have shown 
the LRFS measurements in the figure corresponding to 1024 consecutive periods 
and increasing the starting point by fifty periods. The peak frequency in the 
time average LRFS is given as $f_p$ = 0.408$\pm$0.001 cycles/$P$. The FWHM of 
the peak frequency is 0.003 cycles/$P$ and the corresponding $S$ = 337.6, which
is the highest in our list. However, the FWHM is still a factor of two to three
larger than in certain individual realizations of the LRFS. The drifting 
periodicity of the pulsar is $P_3$ = 2.452$\pm$0.006 $P$ which is consistent 
with the measurements of \citet{bas16}.

In figure \ref{fig_J1720}, right plot, the change in the amplitude of the 
frequency peak, as well as the associated phase, with the pulse longitude
are shown. The amplitudes show the widest spread across the mean for this 
pulsar. This is most likely because the single pulses have the weakest signal 
to noise. In this context the high ordering of the drifting is remarkable and 
is seen despite the comparative weakness of the pulsar signal. The phases show 
a large variations increasing steadily from around -500\degr at the leading 
edge of the pulse window to about +500\degr towards the trailing edge. The 
phase changes are not of linear nature but show a relative flattening towards 
the trailing edge. However, the variations are different from PSR J0820$-$1350 
where the phases show more flattening towards both the leading and trailing 
edge.

\section{Discussion}
The most prominent model of subpulse drifting has been proposed by 
\citet{rud75} who considered an aligned rotator. The sparking discharges in 
the IAR lagged behind co-rotation since the plasma density is less than the 
co-rotation density resulting in the drifting pattern. Alternatively, 
there are other existing models of non-aligned and non-corotating 
magnetospheres where the subpulse drifting phenomenon can also to be explored 
\citep{mel13,yue16}. The validity of the rotating subbeams around the magnetic 
axis in the form of `carousel' model in non-aligned rotators was assumed by 
\citet{gil00,des01} and subsequent analysis of subpulse drifting was based on 
this model. The different phase behaviours seen in the population are 
associated with different line of sight cuts of the emission beam. In recent 
works it has been suggested that the carousel model is physically inconsistent 
with the pulsar electrodynamics \citep{bas16,mit17}. In non-aligned pulsars the
lagging behind co-rotation imply that the sparks should be moving around the 
rotation axis and not the magnetic axis as postulated in the carousel model. In
addition many models \citep{gil03} require the presence of highly non-dipolar 
fields in the IAR. The radius of curvature of the field lines in the IAR should
be much higher than the dipolar case for the primary plasma to reach sufficient 
energies for radio emission \citep[see][for a review of the conditions in the 
IAR leading to radio emission in pulsars]{mit17b}. The presence of non-dipolar 
fields in the IAR would considerably distort the magnetic field lines in the 
gap and likely displace the polar cap from its dipolar location. Models for 
such fields exist in the literature \citep{gil02} where the non-dipolar field 
have been approximated as a star centered dipole coupled with dipoles, with 
much smaller dipole moments, on the stellar crust near the IAR. One direct 
consequence of these conditions is that the spark trajectories in the IAR will 
be distorted. The phase variations associated with subpulse drifting as shown 
in our measurements deviate from linear variations across the pulse window. 
This is likely another indication of the presence of non-dipolar magnetic 
fields in the IAR. However, no such study exist in the literature where 
subpulses lagging behind corotation and moving in non-dipolar magnetic fields 
have been simulated to show the nonlinear phase variations. In certain 
cases the deviations of the phase variations from linear trend can be explained
via line of sight curvature \citep{edw03}.

Another aspect of subpulse drifting that has been particularly challenging to
understand is the effect of bi-drifting seen in only three pulsars, including
J1034$-$3224. The reversal of phase changes in adjacent profile components 
cannot be explained either from a simple carousel rotation of subpulses around
the magnetic axis as well as the sparks lagging behind corotation around the 
rotation axis. To understand this effect several exotic models for emission has
been suggested. \citet{qia04a} proposed the presence of an Inner Annular Gap in
addition to the IAR. In the two gaps the drifting is assumed to be different 
which accounts for the bi-drifting phenomenon. In recent works it has been 
suggested that the polar cap associated with the IAR is highly distorted and
is elliptical in nature and located further way from the dipolar polar cap. In
such cases the bi-drifting effect has been reproduced using the carousel model
and and the associated line of sight traversing a complex path through the IAR 
\citep{sza17,wri17}. It should be noted that in all three pulsars where 
bi-drifting have been observed the profiles are very wide encompassing more 
than 100\degr~in longitude. This is indicative of the pulsar geometry where the
angle between the rotation and magnetic axis should be small. Another promising
model for the bi-drifting which has not been explored yet is the lagging behind
corotation in the presence of highly non-dipolar magnetic fields in the IAR.

Finally, our studies also show that in the majority of cases the drifting 
feature is not seen as a narrow peak but shows certain width which further 
evolves with time. It is possible that the $P_3$ is not sharply defined but 
jitters around a mean value. It should be noted that the subpulses are 
themselves averaged over thousands of sparks. The origin of any variation on 
this average effect is not clear to us and requires more detailed modelling of 
the physical conditions in the IAR. In this scenario the sharp peak associated 
with the pulsar J1720$-$2933 is remarkable since it shows highly ordered 
states. However, even in this pulsar we have shown the peak to jitter around 
the mean value in longer duration observations. The truly baffling observation
for us remains the two frequency peaks seen in the pulsar J1555$-$3134 which 
our analysis show are not harmonically related. If we relate $P_3$ to the 
sparking process in the IAR then $P_3$ = $d/v_d$, where $d$ is the average 
separation between two consecutive sparks and $v_d$ the drift velocity of the 
spark. In the IAR $v_d$ can be further expressed as $v_d = (\Delta E/B)c$, 
where $\Delta E$ is the change in the electric field in the gap during the 
sparking process, $B$ the magnetic field in the IAR and $c$ the speed of light.
The variables in the above expression are $d$ and $\Delta E$. The average 
separation between sparks is governed primarily by the mean free path of 
$\gamma$-ray photons in the IAR \citep{rud75}. In a separate work 
\citet{van12} has suggested the drift velocity to depend on the variation of 
the accelerating potential across the polar cap. However, there are no 
provisions in the above models to explain the jittering around the mean 
frequency with timescales spanning several hundred periods. Most of these 
models deal with steady state configuration with only spatial variations across
the IAR. Such requirements for the time variations in the IAR has also been 
noted by \citet{bas17,mit17}, in order to explain different phenomenon like
periodic nulling or mode changing in certain pulsars. It was suggested in these
works that the pair production process in the IAR is periodically or 
quasi-periodically affected by external triggering mechanism at timescales 
ranging from minutes to hours. This can also be relevant for the jittering seen
in the subpulse drifting peaks. There are also no provisions to accommodate 
multiple realizations of either $d$ or $\Delta E$ to coexist simultaneously or 
switch continuously between two states at short intervals. The presence of two 
frequency peaks would require new understanding of the physical processes in 
the pulsar magnetosphere. Alternative models for drifting subpulses like the 
ones involving surface oscillations can be explored to investigate the multiple
simultaneous peaks in the fluctuation spectra \citep{ros08}.

\section{Summary}
\noindent
In this work we have carried out detailed measurements of drifting properties 
in four long period pulsars which show very different behaviours. In three
pulsars J0820$-$1350, J1555$-$3134 and J1720$-$2933 the phase variations are 
monotonically increasing from leading to the trailing edge, however these 
variations are not linear. On the other hand the pulsar J1034$-$3224 show the 
rare phenomenon of bi-drifting with reversals in the direction of phase 
variation in every alternate profile component. The pulsar J1555$-$3134 show 
the presence of two different drifting frequencies at the same time which are 
not harmonically related. The peak frequency associated with drifting usually 
show a jitter around a mean value. This is also true for the most ordered drift
pattern in PSR J1720$-$2933 which shows small scale variations in peak 
frequency with time. The diversity of subpulse drifting reported here is 
particularly challenging for the different models predicting the physical 
conditions in the pulsar magnetosphere. The temporal fluctuations of the 
drifting frequencies, the shape of the phase variations and the presence of the
two peaks in the pulsar J1555$-$3134 point towards unexplored physical 
processes that affect the emission mechanism. The physical processes 
responsible for the plasma generation take place at timescales of hundreds of 
nanoseconds to several tens of microseconds \citep{rud75}. Despite the temporal
fluctuations of the peak frequency from the mean value the subpulse drifting 
exist in an underlying steady state throughout these long observing runs. This 
demonstrates the pulsar magnetosphere to exist in a steady emission state over 
long timescales.

\section*{Acknowledgments}
We thank the referee for the detailed comments and especially for the help in 
clarifying our arguments for the two peaks in PSR J1555$-$3134 not being 
harmonically related. We thank the staff of the GMRT who have made these 
observations possible. The GMRT is run by the National Centre for Radio 
Astrophysics of the Tata Institute of Fundamental Research.

\end{document}